\documentclass{aa}

\usepackage[varg]{txfonts}
\usepackage{graphicx}
\usepackage{amsmath}
\usepackage{amssymb}
\usepackage{footmisc}
\usepackage{lscape}
\usepackage{color}
\usepackage{longtable}
\usepackage{natbib,twoopt}
\bibpunct{(}{)}{;}{a}{}{,} 

\begin{document} 

\title{Colour variations in the GRB\,120327A afterglow}

\author{
A. Melandri \inst{1},
S. Covino \inst{1},
E. Zaninoni  \inst{2},
S. Campana \inst{1},
J. Bolmer \inst{3},
B. E. Cobb \inst{4},
J. Gorosabel \inst{5}\fnmsep\thanks{Deceased},
J.-W. Kim \inst{6,27},
P. Kuin \inst{7},
D. Kuroda \inst{8},
D. Malesani \inst{9},
C. G. Mundell \inst{10,11},
F. Nappo \inst{1},
B. Sbarufatti \inst{12},
R. J. Smith \inst{10},
I. A. Steele \inst{10},
M. Topinka \inst{13},
A. S. Trotter \inst{14,15},
F. J. Virgili \inst{10},
M. G. Bernardini \inst{16,1},
P. D'Avanzo \inst{1},
V. D'Elia \inst{17,18},
D. Fugazza \inst{1},
G. Ghirlanda \inst{1},
A. Gomboc \inst{19},
J. Greiner \inst{3},
C. Guidorzi \inst{20},
J. B. Haislip \inst{14},
H. Hanayama \inst{21},
L. Hanlon \inst{22},
M. Im \inst{6},
K. M. Ivarsen \inst{14},
J. Japelj \inst{23},
M. Jel\'{i}nek \inst{24},
N. Kawai \inst{8},
S. Kobayashi \inst{10},
D. Kopac \inst{23},
A. P. LaCluyz\'e \inst{14},
A. Martin-Carrillo \inst{22},
D. Murphy \inst{22},
D. E. Reichart \inst{14},
R. Salvaterra \inst{25},
O. S. Salafia \inst{1},
G. Tagliaferri \inst{1},
S. D. Vergani \inst{26}
}

\institute{
INAF / Osservatorio Astronomico di Brera, via E. Bianchi 36, I-23807 Merate (LC), Italy
\and
ICRANet-Rio, Centro Brasileiro de Pesquisas F\'isicas, Rua Dr. Xavier Sigaud 150, Rio de Janeiro, RJ, 22290-180, Brazil
\and
Max-Planck-Institut f\"{u}r extraterrestrische Physik, Giessenbachstrasse 1, D-85748 Garching, Germany
\and
Department of Physics, The George Washington University, Washington, D.C. 20052, USA
\and
Instituto de Astrof\'{\i}sica de Andaluc\'{\i}a (IAA-CSIC), Glorieta de la Astronom\'{\i}a s/n,18008, Granada, Spain
\and 
CEOU / Astronomy Program, Dept. of Physics \& Astronomy, Seoul National University, Seoul, South Korea
\and
Mullard Space Science Laboratory, University College London, Holmbury St Mary, Dorking, Surrey RH5 6NT, UK
\and
Okayama Astrophysical Observatory, National Astronomical Observatory of Japan, Asakuchi, Okayama 719-0232, Japan
\and
Dark Cosmology Centre, Niels Bohr Institute, University of Copenhagen, Juliane Maries Vej 30, 2100 Copenhagen, Denmark
\and
Astrophysics Research Institute, Liverpool JMU, IC2, Liverpool Science Park, 146 Brownlow Hill, Liverpool L3 5RF, UK 
\and
Department of Physics, University of Bath, Claverton Down, Bath, BA2 7AY, UK
\and
Department of Astronomy and Astrophysics, Pennsylvania State University, University Park, PA 16802, USA
\and
Dublin Institute for Advanced Studies, 31 Fitzwilliam Place, Dublin 2, Ireland
\and 
Skynet Robotic Telescope Network, Dep. of Physics and Astronomy, Univ. of North Carolina, Chapel Hill, NC 27599-3255, USA
\and
Department of Physics, North Carolina A\&T State University, Greensboro, NC 27411, USA
\and
Laboratoire Univers et Particules de Montpellier, Universit\'{e} Montpellier, Place Eug\'{e}ne Bataillon, F-34095, Montpellier, France
\and
INAF / Osservatorio Astronomico di Roma, via Frascati 33, I-00078 Monteporzio Catone (Roma), Italy
\and 
ASI Science Data Centre, Via del Politecnico snc, I-00133 Roma, Italy
\and
Centre for Astrophysics and Cosmology, University of Nova Gorica, Vipavska 11c, 5270 Ajdov\v s\v cina, Slovenia
\and
Dipartimento di Fisica, Universit\'{a} di Ferrara, via Saragat 1, I-44100 Ferrara, Italy
\and
Ishigakijima Astronomical Observatory, NAO of Japan, 1024-1, Arakawa, Ishigaki, Okinawa 907-0024, Japan
\and
University College Dublin, School of Physics UCD, Belfield, Dublin 4, Ireland
\and
Faculty of Mathematics and Physics, University of Ljubljana, Jadranska 19, 1000 Ljubljana, Slovenia
\and
Astronomical Institute, Czech Academy of Sciences, (ASU CAS), Ond\v{r}ejov, Czech Republic
\and
INAF / IASF Milano, via E. Bassini 15, I-20133 Milano, Italy
\and
GEPI, Observatoire de Paris, CNRS, Univ. Paris Diderot, 5 place Jules Janssen, F-92190 Meudon, France
\and
Korea Astronomy and Space Science Institute, Daejeon 34055, Korea
}

\offprints{andrea.melandri@brera.inaf.it}
\date{Received ; accepted }

\abstract
   {}
   {We present a comprehensive temporal and spectral analysis of the long {\it Swift} GRB\,120327A afterglow data to investigate the possible causes of the observed early time colour variations.}
   {We collected data from various instruments/telescopes in different bands (X-rays, ultra-violet, optical and near-infrared) and determined the shapes of the afterglow early-time light curves. We studied the overall temporal behaviour and the spectral energy distributions from early to late times.}
   {The ultra-violet, optical, and near-infrared light curves can be modelled with a single power-law component between 200 and 2 $\times$ 10$^{4}$~s after the burst event. The X-ray light curve shows a canonical steep-shallow-steep behaviour, typical of long gamma-ray bursts. At early times a colour variation is observed in the ultra-violet/optical bands, while at very late times a hint of a re-brightening is visible. The observed early time colour change can be explained as a variation in the intrinsic optical spectral index, rather than an evolution of the optical extinction.}
   {}


\titlerunning{Colour variations in GRB\,120327A afterglow}
\authorrunning{A. Melandri et al.}

\maketitle


\section{Introduction}

The early afterglow is one of most interesting emission stages of Gamma-Ray Bursts (GRBs). In a few tens of seconds the afterglow emission (X-ray, ultra-violet, optical, infrared and radio) begins to dominate over the fading prompt (gamma-ray) emission, and the time-scale and intensity of the phenomenon offer powerful diagnostics about the physical processes within the outflow and about the environment of the progenitor \citep[e.g.][]{Ves06,Mol07,Mel08,Lia13,Zan13}.

Of particular interest are those few cases where a colour variation during the early optical afterglow has been singled out \citep{Nys06,Mor14}. GRB\,120327A is one of such a few events. In general, early time colour variation can be the signature of rather different phenomena, e.g. the passage of a break frequency of the optical spectrum through the optical bands \citep[e.g.][]{Fil11}, variation of the optical extinction due to either dust photo-destruction \citep{Mor14} or the outflow progression through a wind-shaped environment \citep{Ryk04}, and also the superposition of different emission stages, such as forward and reverse shock \citep{KoZh03} in the context of the ``fireball" model \citep{Pir04}.

In this paper we report and discuss the observations of the long GRB\,120327A, concentrating on the early time colour evolution of the optical light curve. Data are reported in Sect.\,\ref{sec:data}. The results are presented in Sect.\,\ref{sec:results}. The possible interpretative scenarios are discussed in Sect.\,\ref{sec:meth} and conclusions are drawn in Sect.\,\ref{sec:con}. The respective temporal and spectral decay indices $\alpha$ and $\beta$ are defined by $f_{\nu} (t) \propto t^{-\alpha} \nu^{-\beta}$, and unless stated otherwise, all errors are reported at 1$\sigma$.


\section{Observations}
\label{sec:data}

GRB\,120327A was discovered by the {\it Swift} satellite \citep{Geh04} on March 27, 2012, at 02:55:16 UT \citep{Sba12}. The bright X-ray and optical counterparts of this long gamma-ray burst \citep[$T_{\rm 90} \sim 63$~s;][]{krimm} were observed by the X-ray Telescope (XRT) and the Ultraviolet and Optical Telescope (UVOT). In the UV it was at $U \sim 18$ a few minutes after the prompt event \citep{Kui12}. A redshift was quickly measured by \citet{PeTa12} and \citet{Kru12} by absorption lines at $z = 2.813$. Several ground-based facilities observed the field and detected the counterpart in the optical and near-infrared wavelengths. The afterglow was also detected at 34\,GHz with a flux density of about 0.7\,mJy \citep{Han12}.

In this paper we retrieved and analysed XRT and UVOT data, together with REM telescope \citep{Zer01, Cov04} near-infrared (NIR) data, IAC80, CTIO, BOOTES, PROMPT, Watcher,  CQUEAN/2.1m Otto-Struve telescope \citep{park12}, GROND \citep{Grond}, SMARTS, MITSuME, NOT, and Liverpool Telescope \citep[LT;][]{LT04} optical data. For the analysis we included also photometric data from LT-RINGO2 imaging polarimeter \citep{Kopac17}. Data are reduced and analysed following standard procedures. Calibration was obtained by means of secondary standard stars in the field provided by the APASS\footnote{http://www.aavso.org/apass} and 2MASS\footnote{http://www.ipac.caltech.edu/2mass/} catalogues in the optical and NIR bands, respectively. $Rc$ and $Ic$ magnitudes have been obtained from APASS $r$ and $i$ magnitudes by means of suitable transformation equations. Optical SMARTS data are calibrated by means of a Landolt standard star field. ZY 2.1m Otto-Struve telescope magnitudes are calibrated following \citet{Hod09}. Optical and NIR magnitudes, not corrected for the Galactic reddening $E_{\rm B-V} = 0.29$ \citep{SF11}, are reported in Table\,\ref{tab:data} (\textit{Online Material}). These data supersede those published in \citet{Kui12},  \citet{Cov12}, \citet{Gor12}, \citet{LaC12}, \citet{Mee12}, \citet{Im12}, \citet{Cob12}, \citet{Kur12a,Kur12b} and \citet{Smi12}. We also used in our analysis data from \citet{Sud12}.


\section{Results}
\label{sec:results}

\subsection{Light curves}

\begin{figure*}
\centering
\includegraphics[width=\textwidth]{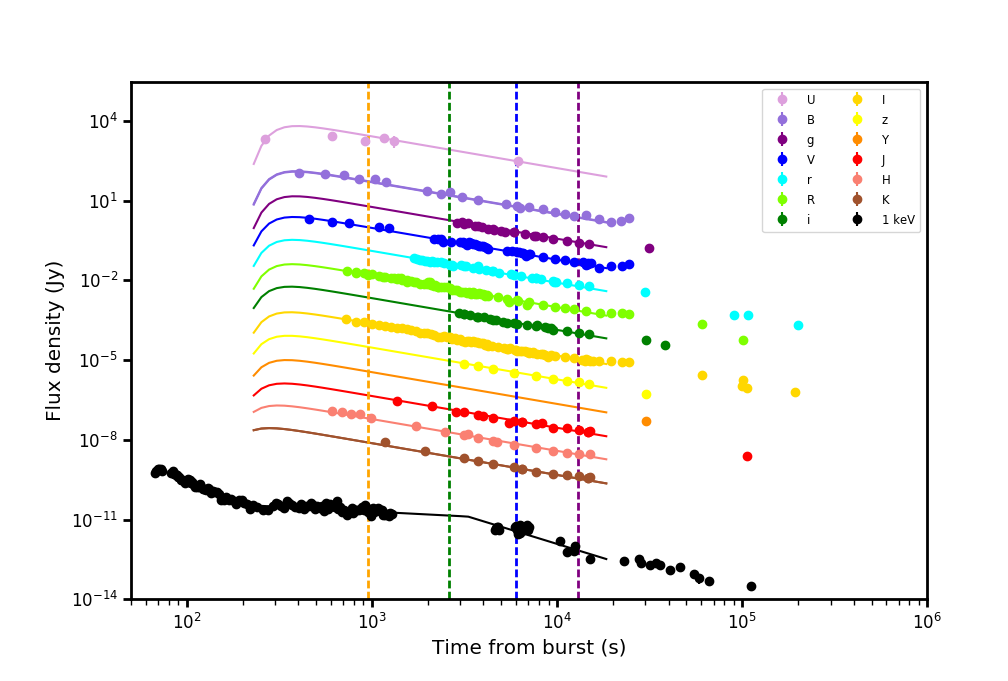}
\caption{Optical (UBgVRrIiz), NIR (YJHK) and X-ray (1 KeV) light curve of GRB\,120327A afterglow. Bands are artificially shifted for clarity (see Table\,\ref{tab:data} for calibrated magnitudes). Solid lines represent the best fit for the optical/NIR bands, assuming a variable optical spectral index $\beta$ and fixed optical extinction. Vertical dashed lines show the T$_{\rm{mid}}$ of the four SEDs described in the main text.}
\label{fig:lc}
\end{figure*}

In Fig.\,\ref{fig:lc} the available optical/NIR data are plotted together with the X-ray light curve at 1\,keV. The X-ray afterglow shows a canonical steep-flat-steep evolution with a considerable variability superposed on the general trend. The initial X-ray decay ($\alpha_{\rm X,1} \sim 3$) is consistent with the tail of the prompt BAT emission. Between $3 \times 10^{2}$ and  $3 \times 10^{3}$ s the light curve flattens to $\alpha_{\rm X,2} \sim 0.4$ and then it becomes steeper to $\alpha_{\rm X,3} \sim 2$ up to the limit of detection. 

\begin{figure}
\centering
\includegraphics[width=\columnwidth,height=8.0cm]{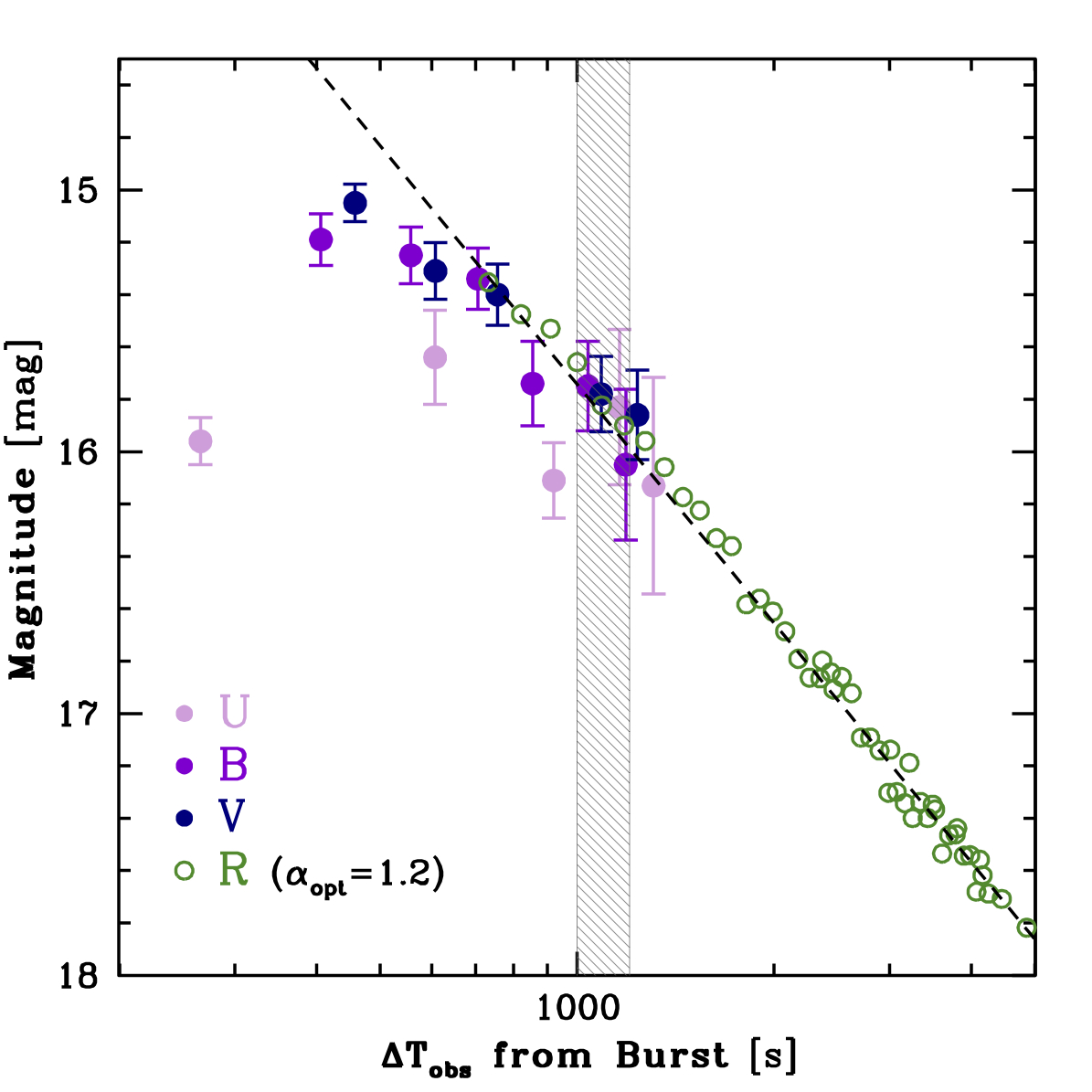}
\caption{Early time $UBV$ magnitudes normalised to the $R$-band light curve at $t \sim 10^{3}$~s (shaded region). The dashed line is the power-law decay index of the late time optical afterglow.}
\label{fig:lc2}
\end{figure}

The optical data start during the X-ray flat phase and show a colour evolution in the optical/NIR afterglow getting redder with time. The evolution is stronger in the bluer bands and more rapid at early-time. We performed a multi-band fit of the UV/optical/NIR data in the time interval [200, 2$\times$10$^{4}$]~seconds (Fig.\,\ref{fig:lc}). Between $10^{3} - 10^{4}$~s the more densely sampled optical bands follow a decay $\alpha_{\rm opt} \sim 1.2$. To visually emphasise the observed colour variation, in Fig.\,\ref{fig:lc2} we normalised all the early time data to the better sampled $R$-band, shifting all the $UBV$ magnitudes in order to have good accordance at the time $t \sim 10^{3}$~s (as shown in Fig.\,\ref{fig:lc} the afterglow temporal behaviour after that time is clearly achromatic in the UV/optical/NIR bands). As can be seen the more we move to bluer filters the larger is the deviation at early times from the estimated afterglow decay index. At late time the optical light curve shows the hint of a re-brightening that evolves from the red to the blue. However, the investigation of this late feature is beyond the purpose of this work.

\subsection{Spectral energy distributions}

For the spectral energy distribution (SED) fit we consider the absorption in the optical and X-ray ranges both locally (i.e., in the GRB host galaxy) and arising from the Milky Way (MW). For the optical band we used the extinction laws given by \cite{Pei92} (Eq. 20 and Table 4 therein) for the MW, the Large Magellanic Cloud (LMC) and Small Magellanic Cloud (SMC). For the X-ray data, we considered the model for the photoelectric cross section per HI-atom units for a given metallicity presented by \cite{Mor83}, assuming solar metallicity \citep[e.g.][]{Cov13,Zan13}.

In addition, \citet{Del14} showed that the SED of GRB\,120327A is characterised by a powerful  Lyman$_\alpha$ emission with $\log N_{\rm H} \sim 22$. This implies that the obtained photometry in the bands including hydrogen lines has to be corrected for the line emission. Moreover, some of our data are obtained with filters that cover spectral ranges bluer than the Lyman$_\alpha$, i.e. about $460$\,nm in the observer frame. Together with the host galaxy extinction we therefore need to consider the absorption due to intergalactic-medium that can only be computed in statistical sense. We followed the recipe proposed by \cite{Jap12}.

\begin{figure}
\centering
\includegraphics[width=\columnwidth,height=7.0cm]{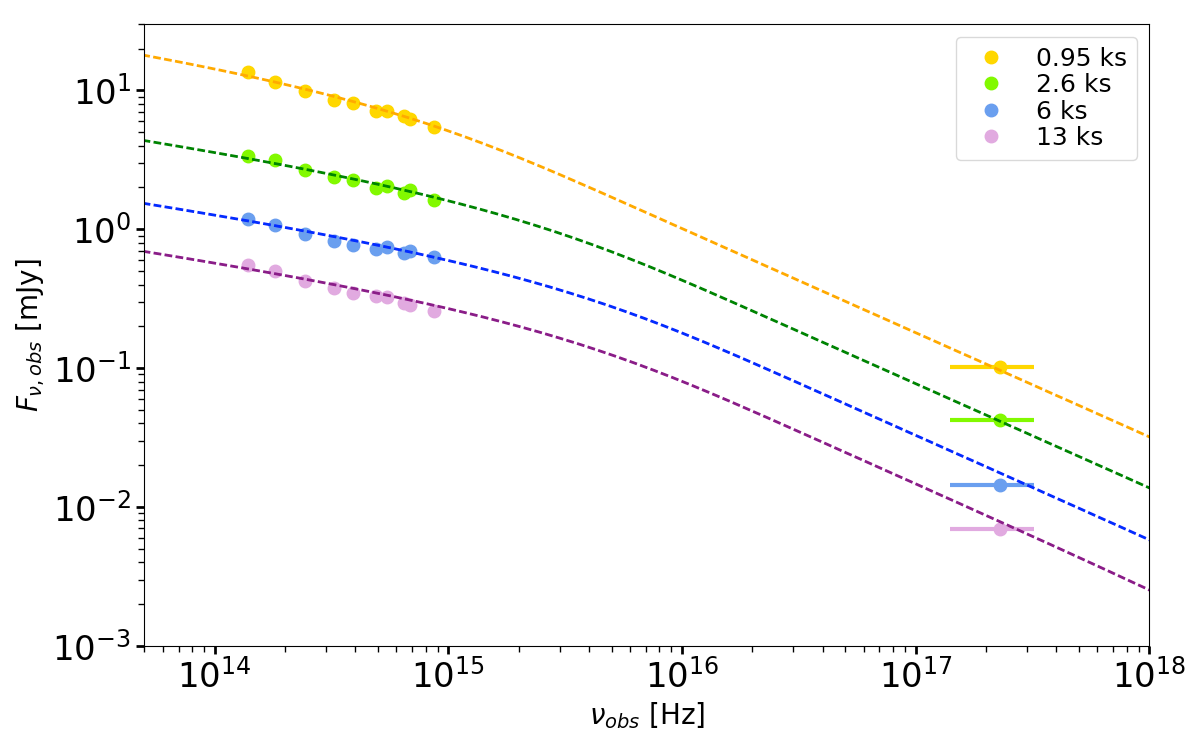}
\caption{The four spectral energy distributions estimated at T$_{\rm{mid}}$. De-reddened optical data and un-absorbed X-ray flux are shown. Dashed lines represent our fitting model.}
\label{fig:SED_all}
\end{figure}

\begin{figure}
\centering
\includegraphics[width=\columnwidth,height=8.0cm]{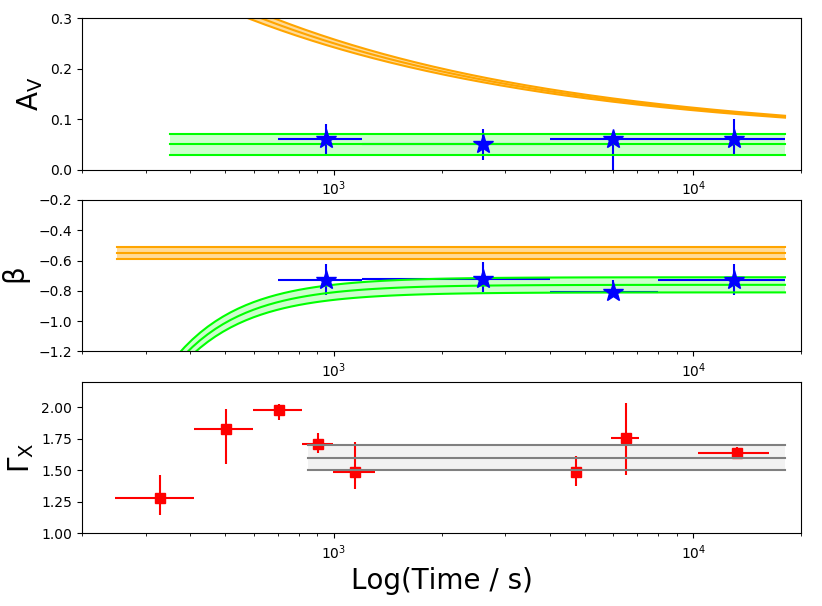}
\caption{Evolution with time of the spectral parameters. Variation of the optical extinction A$_{\rm{V}}$ (\textit{top}) and of the optical spectral index $\beta$ (\textit{center}). \textit{Blue stars} represent the values obtained by four SEDs as reported in Table \ref{tab:SED}. \textit{Green area} refers to the model with A$_{\rm{V}}$ fixed, while \textit{orange area} to the model with $\beta$ fixed. \textit{Red squares} (\textit{bottom}) represent the variation of the X-ray photon index $\Gamma$, that at late times it settles around the value $\sim$1.6 $\pm$ 0.1 (\textit{gray area}).}
\label{fig:SED}
\end{figure}

In Fig. \ref{fig:SED_all} and Fig. \ref{fig:SED} we show the results of the SEDs performed at four different time intervals between 700 and 18000 s after the burst event. The X-ray-to-optical spectral energy distributions are best fitted using a broken power-law\footnote{A valid solution can still be obtained assuming a single power-law with $\beta_{\rm OX} \sim 0.85$. However, since the temporal decays in the X-ray and optical bands are different at the time of each SED, this does not seem to be the more plausible solution.}. 

We fit the data considering two possible models (Fig. \ref{fig:SED}):

\begin{itemize}
\item Variable optical extinction as A$_{\rm{V}}$ = A$_{\rm{V0}} + k \times  t^{0.5}$, with A$_{\rm{V0}} = 0.04 \pm 0.02$, $k = 6.02 \pm 0.25$ and $t$ the time in seconds from the burst onset\footnote{This function is an upgrade of the optical flux attenuation function described by \cite{Ryk04}, assuming the bulk Lorentz factor $\Gamma$ of the emitting shell free to vary.}; the optical spectral index is constant $\beta = 0.55^{+0.05}_{-0.04}$ (\textit{orange area}). 
\item Variable optical spectral index as $\beta = \beta_0 + k t^\alpha$, with $\beta_0 = 0.76 \pm 0.05$, $k=-5.2 \rm{e} 5$, $\alpha= -2.34^{+0.23}_{-0.26}$, and $t$ the time in seconds from the burst onset; the optical extinction is constant A$_{\rm{V}}= 0.05\pm0.02$ (\textit{green area}).
\end{itemize}

The data obtained performing the four SEDs (Fig. \ref{fig:SED_all}) are in accordance with the second (empirical) model, that is a variable optical spectral index and a constant optical extinction ($\chi^2_{\rm red} = 1.18$ for 333 d.o.f.). As said before, the X-ray light curve shows the plateau phase between 200 and $\sim$3000 s. In this epoch the photon index $\Gamma_{\rm X}$ rises from $\sim$ 1.2 at $\sim$ 300 s up to $\sim$ 2.0 at $\sim$ 700 s and then seems to decrease until it sets at the value of $\sim$ 1.6 $\pm$ 0.1 (Fig. \ref{fig:SED}. bottom panel). This is similar to the variation of the optical spectral index, as reported in Table \ref{tab:SED} and fitted in Fig. \ref{fig:SED} (mid panel), strengthening the validity of our fit with a broken power-law function.

\begin{table}
\centering
\caption{Fit parameters of the optical spectral energy distribution at different epochs. $^{\rm{(a)}}$ Without U filter.}
\begin{tabular}{ccccc}
\hline
\hline
T$_{\rm{min}}$ (s) & T$_{\rm{max}}$ (s) & T$_{\rm{mid}}$ (s) &A$_{\rm{V}}$ & $\beta$\\
\hline
700$^{\rm{(a)}}$    & 1200   & 950 & 0.06$^{+0.03}_{-0.03}$ & 0.73$^{+0.11}_{-0.10}$\\
1200  & 4000   & 2600 & 0.05$\pm$0.03 & 0.72$^{+0.11}_{-0.09}$\\
4000  & 8000   & 6000 & $\le 0.05$ & 0.81$^{+0.08}_{-0.02}$\\
8000  & 18000 & 13000 & 0.06$^{+0.04}_{-0.03}$ & 0.73$^{+0.11}_{-0.10}$\\
\hline
\hline
\end{tabular}
\label{tab:SED}
\end{table}

\section{Discussion}
\label{sec:meth}

\subsection{Early time colour variation}

In the context of the standard fireball model \citep{Sar98, ChLi00} the GRB afterglow is related to the synchrotron emission from a decelerating relativistic shell that slows into an external medium. The observed colour variation in the optical light curve at early time (Fig.\,\ref{fig:lc} and Fig.\,\ref{fig:lc2}) could arise from different scenarios:

\begin{enumerate}[(a)]

\item the color evolution could be caused by the {\it passage of a break frequency}. Considering that the spectral index $\beta$ becomes shallower, the passage of the cooling frequency $\nu_{\rm c} \propto t^{1/2}$ in the slow cooling wind medium is only the possible case in this scenario, otherwise $\beta$ becomes steeper or it changes sign \citep{ChLi00}. The passage of $\nu_{\rm c}$ is expected to cause the spectral index change $\Delta \beta = 1/2$, simultaneously with the decay index change $\Delta \alpha = 1/4$ in the light curve \citep{Sar98}. However, the observed change in the spectral index is smaller $\Delta \beta \sim 0.2$ (i.e. the difference in $\beta$ between the 1st epoch and the later epochs), and no temporal breaks seem to be associated with this spectral change. Moreover, our spectral energy distributions analysis seems to indicate that $\nu_{\rm opt} < \nu_{\rm c} < \nu_{\rm X}$ at the time of each SED (Fig. \ref{fig:SED_all}). Therefore the passage of a break frequency can not explain the observations well, and the scenario can be discarded;
~\\
\item the colour change could be simply caused by a {\it variation of the spectral index $\beta$}. Following \citet{Mor14}, we fit the SEDs letting $\beta$ free to vary and fixing the dust parameters to the values obtained at late times. Therefore, the colour change can be modelled like a variation in the intrinsic spectral index $\beta$, as $F_{\nu_2} = F_{\nu_1} (\nu_2/\nu_1)^{\Delta\beta_{12}}$ \citep{Per10}. Indeed, as can be seen in Fig. \ref{fig:SED} (middle panel), this possibility could explain the observed data. From our fit, the largest variation of the spectral index is expected at the very early times (green area), where it goes from $\beta \sim -1.4$ at the time of our first UV detection ($\sim 3 \times 10^{2}$~s) up to $\beta \sim -0.7$ at t$\sim 10^{3}$~s (when in fact the observed temporal behaviour becomes achromatic);
~\\
\item another possible source of the colour variation could be a {\it change of the optical extinction because of the dust destruction} by the jet within 10 - 30 pc \citep{Dra79,Wax00,Fru01,Per02,Dra02}. This is supported by the fact that long GRBs have massive star progenitors and explode in dusty environments  \citep{Mor14}. In this case, both the extinction $A_V$ and the reddening $R_V$ are expected to change \citep[e.g.][]{Per03}. Therefore, we fitted the optical/X-ray SEDs letting $A_V$ and $\beta$ free to change, since the dust absorption observed at early times can have different signatures than those at late time \citep{Mor14}. For GRB\,120327A we do not see a clear variation of the extinction in our spectral analysis. However, we find an excess of the X-ray absorbing column density at early time (N$_{\rm H} \sim 1.6 \times 10^{22}$~cm$^{-2}$) that disappears after $\sim$ 150~s from the burst onset, becoming consistent with the Galactic value \citep{Wil13}. Assuming the dust-to-gas relation reported by \citet{Cov13} this would correspond to an $A_{\rm V} \sim$ 1 mag, implying some sort of extinction variation at very early times. However, the dust destruction scenario seems a contrived explanation.

~\\

\end{enumerate}


\section{Summary and conclusions}
\label{sec:con}

We analysed the temporal and spectral properties of GRB\,120327A. The multi-band fit has highlighted the presence of a colour variation from early to late time that cannot be easily explained with the passage of a break frequency through the optical/NIR bands, either with the assumption of an homogenous or wind-like surrounding medium. However, in the fast-cooling case with a wind-like medium, small changes of the spectral index could be the result of the theoretical curvature of the spectrum \citep{Gra02}.

No evidence for a change in the absorption $A_V$ (we find an average $A_V= 0.05 \pm 0.02$ for the dust content of the host galaxy), that could explain the observed colour variation, is clearly seen in our data. Although we cannot exclude completely the dust photo-destruction scenario, the variation of the spectral index $\beta$ seems to be the favoured explanation, reproducing the observed data quite well. Such a variation could be the result of small changes of the microphysical parameters ($p$, $\epsilon_{\rm e}$, and $\epsilon_{\rm B}$) from early to late times. In particular, a small variation of electron spectral index (of the order of $\sim 0.4$~dex) would reproduce the observed $\Delta  \beta$.

~\\


\begin{acknowledgements}
AM, SCo, SCa, BS, PDA, and GT acknowledge support from the ASI grant I/004/11/3. EZ acknowledges the support by the International Cooperation Program CAPES-ICRANet financed by CAPES - Brazilian Federal Agency for Support and Evaluation of Graduate Education within the Ministry of Education of Brazil. DM acknowledges support from the Instrument center for Danish Astrophysics (IDA). CGM acknowledges support from the Royal Society, the Wolfson Foundation and the Science and Technology Facilities Council. LH acknowledges support from SFI (07-RFP-PHYF295, 11/RFP.1/AST/3188) $\&$ the EU-FP7/GLORIA (grant no. 283783). MI and JWK acknowledge the support from the National Research Foundation of Korea grant no. 2017R1A3A3001362 and no. 2016R1D1A1B03934815. This work has been supported by ASI grant I/004/11/0 and by PRIN-MIUR 2009 grants. This research was made possible through the use of the AAVSO Photometric All-Sky Survey (APASS), funded by the Robert Martin Ayers Sciences Fund. Partly based on observations made with the Nordic Optical Telescope (program 46-003, PI Jakobsson), operated by the Nordic Optical Telescope Scientific Association at the Observatorio del Roque de los Muchachos, La Palma, Spain, of the Instituto de Astrof\'isica de Canarias.
\end{acknowledgements}

\bibliographystyle{aa} 

\Online
\onecolumn
\begin{longtab}
\footnotesize{
\begin{longtable}{ccccc|ccccc}
\caption{Photometric data used in this paper. Magnitudes are in the Vega system unless for $griz$ filters that are in the AB system, and they are all not corrected for Galactic absorption. Errors are at 1$\sigma$ level.}\label{tab:data} \\
\hline\hline
t-t$_0$ (s) & Exp. (s) & Mag & Band & Telescope & t-t$_0$ (s) & Exp. (s) & Mag & Band & Telescope\\
\hline
\endfirsthead
\caption{Continued.}\\
\hline\hline
t-t$_0$ (s) & Exp. (s) & Mag & Band & Telescope & t-t$_0$ (s) & Exp. (s) & Mag & Band & Telescope\\
\hline
\endhead
\hline
30224 & 360 & 19.10$\pm$0.02 & z & CQUEAN & 266  & 125 & 17.91$\pm$0.09 & u & UVOT \\ 
3126  & 177 & 16.76$\pm$0.03 & z & GROND & 606  & 39  & 17.59$\pm$0.20 & u & UVOT \\
3728  & 348 & 16.98$\pm$0.03 & z & GROND & 921  & 102  & 18.06$\pm$0.18 & u & UVOT \\
4508  & 347 & 17.27$\pm$0.03 & z & GROND & 1162 & 10  & 17.78$\pm$0.33 & u & UVOT \\
5811  & 871 & 17.61$\pm$0.03 & z & GROND & 1308 & 9   & 18.08$\pm$0.46 & u & UVOT \\
7651  & 871 & 17.92$\pm$0.03 & z & GROND & 6166 & 48  & 20.01$\pm$0.44 & u & UVOT \\
9479  & 866 & 18.20$\pm$0.03 & z & GROND &  406  & 10 & 16.99$\pm$0.11 & b & UVOT \\
11305 & 873 & 18.35$\pm$0.03 & z & GROND &  557  & 10 & 17.05$\pm$0.12 & b & UVOT \\
13134 & 867 & 18.49$\pm$0.03 & z & GROND &  706  & 10 & 17.14$\pm$0.13 & b & UVOT \\
14959 & 866 & 18.61$\pm$0.03 & z & GROND & 855  & 10 & 17.54$\pm$0.18 & b & UVOT \\
30376 & 360 & 18.95$\pm$0.05 & Y & CQUEAN & 1038 & 10 & 17.55$\pm$0.19 & b & UVOT \\
1366 &20  &14.34$\pm$0.04 & J & REM & 1186 & 10 & 17.85$\pm$0.32 & b & UVOT \\
2112 &246 &14.83$\pm$0.06 & J & REM & 6268 & 48 & 20.26$\pm$0.39 & b & UVOT \\
2859 &150 &15.41$\pm$0.09 & J & REM & 458  & 10 & 15.65$\pm$0.08 & v & UVOT \\
3982 &300 &15.77$\pm$0.09 & J & REM & 608  & 10 & 15.91$\pm$0.12 & v & UVOT \\
5470 &300 &16.39$\pm$0.15 & J & REM & 756  & 10 & 16.00$\pm$0.13 & v & UVOT \\
8341 &300 &16.43$\pm$0.13 & J & REM & 1088 & 10 & 16.38$\pm$0.16 & v & UVOT \\
6452   & 720  & 16.31$\pm$0.07 & J &SMARTS & 1236 & 10 & 16.46$\pm$0.19 & v & UVOT \\
14628  & 720  & 17.27$\pm$0.08 & J &SMARTS &  6762 & 48 & 19.10$\pm$0.40 & v & UVOT \\
106377 & 1800 & 19.56$\pm$0.14 & J &SMARTS & 2177  & 10 & 17.53$\pm$0.08 & V & PROMPT1\\
3149  & 200 & 16.26$\pm$0.05 & J & GROND & 2263  & 10 & 17.56$\pm$0.09 & V & PROMPT1\\ 
3756  & 376 & 16.57$\pm$0.05 & J & GROND & 2349  & 10 & 17.55$\pm$0.08 & V & PROMPT1\\
4535  & 374 & 16.82$\pm$0.05 & J & GROND & 2435  & 10 & 17.81$\pm$0.11 & V & PROMPT1\\
5844  & 895 & 17.14$\pm$0.05 & J & GROND & 2659  & 10 & 17.76$\pm$0.08 & V & PROMPT1\\
7674  & 895 & 17.42$\pm$0.05 & J & GROND & 3003  & 10 & 17.85$\pm$0.08 & V & PROMPT1\\
9503  & 890 & 17.75$\pm$0.05 & J & GROND & 3090  & 10 & 17.84$\pm$0.07 & V & PROMPT1\\
11329 & 896 & 17.83$\pm$0.05 & J & GROND & 3158  & 10 & 17.85$\pm$0.12 & V & PROMPT1\\
13158 & 891 & 17.95$\pm$0.05 & J & GROND & 3256  & 10 & 18.05$\pm$0.08 & V & PROMPT1\\
14982 & 890 & 18.11$\pm$0.05 & J & GROND & 3344  & 10 & 17.80$\pm$0.07 & V & PROMPT1\\
6480  & 360 & 16.31$\pm$0.07 & J & CTIO & 3432  & 10 & 17.93$\pm$0.08 & V & PROMPT1\\
14760 & 360 & 17.27$\pm$0.08 & J & CTIO & 3519  & 10 & 18.01$\pm$0.08 & V & PROMPT1\\
608  &50  &12.32$\pm$0.04 & H & REM & 3607  & 10 & 18.00$\pm$0.08 & V & PROMPT1\\
693  &50  &12.38$\pm$0.04 & H & REM & 3694  & 10 & 18.12$\pm$0.08 & V & PROMPT1\\
775  &50  &12.59$\pm$0.05 & H & REM & 3782  & 10 & 18.15$\pm$0.09 & V & PROMPT1\\
859  &50  &12.59$\pm$0.05 & H & REM & 3898  & 10 & 18.22$\pm$0.09 & V & PROMPT1\\
992  &150 &12.93$\pm$0.06 & H & REM & 3989  & 10 & 18.20$\pm$0.09 & V & PROMPT1\\
1739 &150 &13.68$\pm$0.06 & H & REM & 4076  & 10 & 18.31$\pm$0.10 & V & PROMPT1\\
2485 &150 &14.33$\pm$0.09 & H & REM & 4164  & 10 & 18.35$\pm$0.10 & V & PROMPT1\\
3309 &300 &14.52$\pm$0.09 & H & REM & 4251  & 10 & 18.45$\pm$0.12 & V & PROMPT1\\
4755 &300 &15.22$\pm$0.15 & H & REM & 5348  & 10 & 18.65$\pm$0.04 & V & PROMPT1\\
3149  & 200 & 15.91$\pm$0.05 & H & GROND & 6086  & 10 & 18.73$\pm$0.06 & V & PROMPT1\\
3756  & 376 & 16.24$\pm$0.05 & H & GROND & 7078  & 10 & 18.94$\pm$0.04 & V & PROMPT1\\
4535  & 374 & 16.52$\pm$0.05 & H & GROND & 8413  & 10 & 19.21$\pm$0.05 & V & PROMPT1\\
5844  & 895 & 16.87$\pm$0.05 & H & GROND & 9727  & 10 & 19.42$\pm$0.05 & V & PROMPT1\\
7674  & 895 & 17.17$\pm$0.05 & H & GROND & 11056 & 10 & 19.47$\pm$0.06 & V & PROMPT1\\
9503  & 890 & 17.46$\pm$0.05 & H & GROND & 12387 & 10 & 19.66$\pm$0.07 & V & PROMPT1\\
11329 & 896 & 17.63$\pm$0.05 & H & GROND & 14326 & 10 & 19.94$\pm$0.07 & V & PROMPT1\\
13158 & 891 & 17.74$\pm$0.05 & H & GROND & 16927 & 10 & 20.24$\pm$0.10 & V & PROMPT1\\
14982 & 890 & 17.77$\pm$0.05 & H & GROND & 19515 & 10 & 20.02$\pm$0.08 & V & PROMPT1\\
1179 &150 &12.30 $\pm$ 0.06 & K & REM & 22302 & 10 & 20.02$\pm$0.07 & V & PROMPT1\\
1925 &150 &13.12 $\pm$ 0.10 & K & REM & 24398 & 10 & 19.86$\pm$0.06 & V & PROMPT1\\  
6452  & 720 & 14.78$\pm$0.09 & K & SMARTS & 6480  & 450 & 18.82$\pm$0.03 & V & CTIO\\
14627 & 720 & 15.68$\pm$0.09 & K & SMARTS & 14760 & 450 & 19.87$\pm$0.03 & V & CTIO\\
3149  & 200 & 15.61$\pm$0.06 & K & GROND & 5676  & 225 & 18.66$\pm$0.04 & V & SMARTS \\
3756  & 376 & 15.87$\pm$0.06 & K & GROND & 6255  & 225 & 18.78$\pm$0.04 & V & SMARTS \\
4535  & 374 & 16.18$\pm$0.06 & K & GROND & 6540  & 225 & 18.86$\pm$0.04 & V & SMARTS \\
5843  & 895 & 16.48$\pm$0.06 & K & GROND & 7110  & 225 & 18.94$\pm$0.04 & V & SMARTS \\
7674  & 895 & 16.92$\pm$0.06 & K & GROND & 13854 & 225 & 19.73$\pm$0.04 & V & SMARTS \\
9503  & 890 & 17.12$\pm$0.06 & K & GROND & 14427 & 225 & 19.84$\pm$0.04 & V & SMARTS \\
11329 & 896 & 17.21$\pm$0.06 & K & GROND & 14712 & 225 & 19.79$\pm$0.04 & V & SMARTS \\
13158 & 891 & 17.35$\pm$0.06 & K & GROND & 15282 & 225 & 19.75$\pm$0.04 & V & SMARTS \\
14982 & 890 & 17.41$\pm$0.07 & K & GROND & 1993   & 80 & 18.72$\pm$0.16 & B & PROMPT3\\
6480  & 360 & 14.78$\pm$0.09 & K & CTIO & 2351   & 80 & 18.98$\pm$0.16 & B & PROMPT3\\
14760 & 360 & 15.68$\pm$0.09 & K & CTIO & 2625   & 80 & 18.80$\pm$0.12 & B & PROMPT3\\
656 & 159 & 20.43$\pm$0.59 & uvw1 & UVOT &  3072   & 80 & 19.30$\pm$0.18 & B & PROMPT3\\
3740   & 80 & 19.54$\pm$0.12 & B & PROMPT3 & 3522  & 10 & 17.37$\pm$0.05 &Rc & PROMPT4\\
5283   & 80 & 19.90$\pm$0.18 & B & PROMPT3 & 3611  & 10 & 17.53$\pm$0.06 &Rc & PROMPT4\\
6045   & 80 & 20.09$\pm$0.23 & B & PROMPT3 & 3700  & 10 & 17.46$\pm$0.06 &Rc & PROMPT4\\
7054   & 80 & 20.24$\pm$0.13 & B & PROMPT3 & 3788  & 10 & 17.46$\pm$0.06 &Rc & PROMPT4\\
8383   & 80 & 20.36$\pm$0.13 & B & PROMPT3 & 3898  & 10 & 17.54$\pm$0.07 &Rc & PROMPT4\\
9725   & 80 & 20.63$\pm$0.15 & B & PROMPT3 & 3986  & 10 & 17.54$\pm$0.06 &Rc & PROMPT4\\
11060  & 80 & 20.87$\pm$0.18 & B & PROMPT3 & 4075  & 10 & 17.68$\pm$0.07 &Rc & PROMPT4\\
12385  & 80 & 21.06$\pm$0.26 & B & PROMPT3 & 4163  & 10 & 17.62$\pm$0.06 &Rc & PROMPT4\\
14322  & 80 & 20.99$\pm$0.17 & B & PROMPT3 & 4252  & 10 & 17.69$\pm$0.06 &Rc & PROMPT4\\
16926  & 80 & 21.31$\pm$0.22 & B & PROMPT3 & 5352  & 10 & 17.97$\pm$0.03 &Rc & PROMPT4\\
19510  & 80 & 21.58$\pm$0.26 & B & PROMPT3 & 6090  & 10 & 18.09$\pm$0.04 &Rc & PROMPT4\\
22189  & 80 & 21.51$\pm$0.27 & B & PROMPT3 & 7087  & 10 & 18.28$\pm$0.03 &Rc & PROMPT4\\
24351  & 80 & 21.24$\pm$0.25 & B & PROMPT3 & 8374  & 10 & 18.53$\pm$0.03 &Rc & PROMPT4\\
2873 &  30 & 18.49$\pm$0.10 & g & Liverpool T. & 9735  & 10 & 18.71$\pm$0.03 &Rc & PROMPT4\\
3080 &  30 & 18.44$\pm$0.09 & g & Liverpool T. & 11065 & 10 & 18.85$\pm$0.04 &Rc & PROMPT4\\
3289 &  60 & 18.52$\pm$0.06 & g & Liverpool T. & 12394 & 10 & 18.97$\pm$0.05 &Rc & PROMPT4\\
3598 &  60 & 18.79$\pm$0.07 & g & Liverpool T. & 14347 & 10 & 19.08$\pm$0.04 &Rc & PROMPT4\\
3916 &  60 & 18.90$\pm$0.08 & g & Liverpool T. & 16951 & 10 & 19.26$\pm$0.04 &Rc & PROMPT4\\
4228 &  60 & 19.03$\pm$0.08 & g & Liverpool T. & 19531 & 10 & 19.28$\pm$0.04 &Rc & PROMPT4\\
4588 &  60 & 19.12$\pm$0.08 & g & Liverpool T. & 22304 & 10 & 19.26$\pm$0.04 &Rc & PROMPT4\\
4972 &  60 & 19.25$\pm$0.05 & g & Liverpool T. & 24366 & 10 & 19.36$\pm$0.06 &Rc & PROMPT4\\
5212 & 120 & 19.30$\pm$0.05 & g & Liverpool T. & 100804 & 10 & 21.83$\pm$0.20 &Rc & PROMPT4\\
5865 & 180 & 19.39$\pm$0.03 & g & Liverpool T. & 822 & 2 & 15.50$\pm$0.06 & Rc & Watcher\\
6733 & 120 & 19.54$\pm$0.07 & g & Liverpool T. & 950 & 2 & 15.68$\pm$0.05 & Rc & Watcher\\
7545 & 180 & 19.70$\pm$0.04 & g & Liverpool T. & 1142 & 2 & 15.91$\pm$0.05 & Rc & Watcher\\
8440 & 120 & 19.84$\pm$0.05 & g & Liverpool T. & 1311 & 2 & 16.05$\pm$0.11 & Rc & Watcher\\
31531 & 540 & 20.89$\pm$0.07 & g & CQUEAN  & 1375 & 2 & 16.02$\pm$0.09 & Rc & Watcher\\
3126  & 177 & 18.55$\pm$0.03 & g & GROND & 1439 & 2 & 16.04$\pm$0.11 & Rc & Watcher\\
3728  & 348 & 18.78$\pm$0.03 & g & GROND & 1556 & 2 & 16.42$\pm$0.09 & Rc & Watcher\\
4508  & 347 & 19.06$\pm$0.03 & g & GROND & 1684 & 2 & 16.34$\pm$0.09 & Rc & Watcher\\
5820  & 871 & 19.38$\pm$0.03 & g & GROND & 2006 & 2 & 16.40$\pm$0.12 & Rc & Watcher\\
7651  & 871 & 19.69$\pm$0.03 & g & GROND & 2070 & 2 & 16.38$\pm$0.12 & Rc & Watcher\\
9479  & 866 & 19.99$\pm$0.03 & g & GROND & 2198 & 2 & 16.69$\pm$0.19 & Rc & Watcher\\
11305 & 873 & 20.17$\pm$0.03 & g & GROND & 2262 & 2 & 16.84$\pm$0.24 & Rc & Watcher\\
13134 & 867 & 20.34$\pm$0.03 & g & GROND & 2368 & 30  & 17.15$\pm$0.06 & Rc & Liverpool T.\\
14959 & 866 & 20.48$\pm$0.03 & g & GROND & 2465 & 30  & 17.26$\pm$0.06 & Rc & Liverpool T.\\
732   & 10 & 15.35$\pm$0.01 &Rc & PROMPT4 & 3010 & 30  & 17.49$\pm$0.05 & Rc & Liverpool T.\\
821   & 10 & 15.47$\pm$0.01 &Rc & PROMPT4 & 3218 & 30  & 17.54$\pm$0.05 & Rc & Liverpool T.\\
912   & 10 & 15.53$\pm$0.02 &Rc & PROMPT4 & 3494 & 60  & 17.70$\pm$0.03 & Rc & Liverpool T.\\
1000  & 10 & 15.66$\pm$0.02 &Rc & PROMPT4 & 3806 & 60  & 17.79$\pm$0.03 & Rc & Liverpool T.\\
1091  & 10 & 15.82$\pm$0.02 &Rc & PROMPT4 & 4124 & 60  & 17.91$\pm$0.03 & Rc & Liverpool T.\\
1180  & 10 & 15.90$\pm$0.02 &Rc & PROMPT4 & 4454 & 60  & 18.06$\pm$0.04 & Rc & Liverpool T.\\
1271  & 10 & 15.96$\pm$0.02 &Rc & PROMPT4 & 4862 & 60  & 18.17$\pm$0.03 & Rc & Liverpool T.\\
1360  & 10 & 16.06$\pm$0.02 &Rc & PROMPT4 & 5623 & 120 & 18.31$\pm$0.02 & Rc & Liverpool T.\\
1451  & 10 & 16.17$\pm$0.03 &Rc & PROMPT4 & 6392 & 180 & 18.49$\pm$0.02 & Rc & Liverpool T.\\
1540  & 10 & 16.22$\pm$0.03 &Rc & PROMPT4 & 7296 & 120 & 18.67$\pm$0.03 & Rc & Liverpool T.\\
1632  & 10 & 16.33$\pm$0.03 &Rc & PROMPT4 & 8164 & 180 & 18.79$\pm$0.02 & Rc & Liverpool T.\\
1722  & 10 & 16.36$\pm$0.03 &Rc & PROMPT4 & 9616 & 300 & 19.01$\pm$0.02 & Rc & Liverpool T.\\
1814  & 10 & 16.58$\pm$0.04 &Rc & PROMPT4 & 9927 & 300 & 19.06$\pm$0.02 & Rc & Liverpool T.\\
1901  & 10 & 16.56$\pm$0.03 &Rc & PROMPT4 & 1664 & 90  & 16.67$\pm$0.05 & Rc & Liverpool T.\\
1990  & 10 & 16.61$\pm$0.03 &Rc & PROMPT4 & 7650   & 25 & 18.66$\pm$0.05 & Rc &REM\\
2079  & 10 & 16.69$\pm$0.03 &Rc & PROMPT4 & 107780 & 50 & 22.20$\pm$0.10 & Rc &REM\\    
2177  & 10 & 16.79$\pm$0.04 &Rc & PROMPT4 & 90720  & 1800 & 22.11$\pm$0.15 & Rc & NOT\\
2265  & 10 & 16.86$\pm$0.04 &Rc & PROMPT4 & 201180 & 1800 & 23.06$\pm$0.25 & Rc & NOT\\
2352  & 10 & 16.87$\pm$0.04 &Rc & PROMPT4 & 3438 & 654 & 17.50$\pm$0.12 & Rc & Boote\\
2441  & 10 & 16.84$\pm$0.04 &Rc & PROMPT4 & 4104 & 674 & 17.65$\pm$0.12 & Rc & Boote\\
2537  & 10 & 16.86$\pm$0.04 &Rc & PROMPT4 & 4812 & 748 & 17.82$\pm$0.13 & Rc & Boote\\
2626  & 10 & 16.92$\pm$0.04 &Rc & PROMPT4 & 5520 & 660 & 18.24$\pm$0.21 & Rc & Boote\\
2714  & 10 & 17.09$\pm$0.05 &Rc & PROMPT4 & 6186 & 673 & 18.17$\pm$0.21 & Rc & Boote\\
2803  & 10 & 17.09$\pm$0.05 &Rc & PROMPT4 & 6858 & 656 & 18.59$\pm$0.29 & Rc & Boote\\
2898  & 10 & 17.14$\pm$0.05 &Rc & PROMPT4 & 60769 & 240 & 20.34$\pm$0.15 & Rc & MITSuME\\
2989  & 10 & 17.30$\pm$0.06 &Rc & PROMPT4 & 29859 & 540 & 20.00$\pm$0.02 & r & CQUEAN\\
3078  & 10 & 17.30$\pm$0.05 &Rc & PROMPT4 & 3126  & 177 &17.56$\pm$0.03 & r & GROND\\
3167  & 10 & 17.34$\pm$0.06 &Rc & PROMPT4 & 3728  & 348 &17.56$\pm$0.03 & r & GROND\\
3256  & 10 & 17.40$\pm$0.06 &Rc & PROMPT4 & 4508  & 347 &18.06$\pm$0.03 & r & GROND\\
3345  & 10 & 17.34$\pm$0.05 &Rc & PROMPT4 & 5820  & 871 &18.39$\pm$0.03 & r & GROND\\
3434  & 10 & 17.40$\pm$0.06 &Rc & PROMPT4 & 7651  & 871 &18.69$\pm$0.03 & r & GROND\\
9479  & 866 &18.99$\pm$0.03 & r & GROND & 5561 & 3600 & 17.35$\pm$0.03 & Ic & IAC80\\
11305 & 873 &19.19$\pm$0.03 & r & GROND & 5945 & 3600 & 17.49$\pm$0.03 & Ic & IAC80\\
13134 & 867 &19.32$\pm$0.03 & r & GROND & 6319 & 3600 & 17.57$\pm$0.04 & Ic & IAC80\\
14959 & 866 &19.45$\pm$0.03 & r & GROND & 6700 & 3600 & 17.62$\pm$0.05 & Ic & IAC80\\
724    & 10 & 14.64$\pm$0.01 & Ic & PROMPT5 & 7073 & 3600 & 17.77$\pm$0.04 & Ic & IAC80\\
821    & 10 & 14.84$\pm$0.02 & Ic & PROMPT5 & 7456 & 3600 & 17.71$\pm$0.04 & Ic & IAC80\\
911    & 10 & 14.88$\pm$0.02 & Ic & PROMPT5 & 7829 & 3600 & 17.84$\pm$0.04 & Ic & IAC80\\
1000   & 10 & 15.05$\pm$0.02 & Ic & PROMPT5 & 8201 & 3600 & 17.90$\pm$0.04 & Ic & IAC80\\
1091   & 10 & 15.15$\pm$0.02 & Ic & PROMPT5 & 8569 & 3600 & 17.96$\pm$0.06 & Ic & IAC80\\
1181   & 10 & 15.27$\pm$0.02 & Ic & PROMPT5 & 8942 & 3600 & 18.12$\pm$0.08 & Ic & IAC80\\
1271   & 10 & 15.35$\pm$0.02 & Ic & PROMPT5 & 9324 & 3600 & 17.94$\pm$0.13 & Ic & IAC80\\
1361   & 10 & 15.40$\pm$0.03 & Ic & PROMPT5 & 2941  & 30  & 16.98$\pm$0.04 & Ic& Liverpool T.\\
1451   & 10 & 15.47$\pm$0.03 & Ic & PROMPT5 & 3148  & 30  & 17.04$\pm$0.04 & Ic& Liverpool T.\\
1541   & 10 & 15.51$\pm$0.03 & Ic & PROMPT5 & 3389  & 60  & 17.15$\pm$0.03 & Ic& Liverpool T.\\
1632   & 10 & 15.65$\pm$0.03 & Ic & PROMPT5 & 3698  & 60  & 17.31$\pm$0.03 & Ic& Liverpool T.\\
1723   & 10 & 15.74$\pm$0.04 & Ic & PROMPT5 & 4018  & 60  & 17.39$\pm$0.03 & Ic& Liverpool T.\\
1814   & 10 & 15.96$\pm$0.04 & Ic & PROMPT5 & 4328  & 60  & 17.54$\pm$0.04 & Ic& Liverpool T.\\
1903   & 10 & 15.95$\pm$0.04 & Ic & PROMPT5 & 4690  & 60  & 17.65$\pm$0.03 & Ic& Liverpool T.\\
1993   & 10 & 15.94$\pm$0.04 & Ic & PROMPT5 & 5077  & 60  & 17.85$\pm$0.05 & Ic& Liverpool T.\\
2082   & 10 & 16.02$\pm$0.04 & Ic & PROMPT5 & 5374  & 120 & 17.88$\pm$0.03 & Ic& Liverpool T.\\
2177   & 10 & 16.13$\pm$0.04 & Ic & PROMPT5 & 6105  & 180 & 18.03$\pm$0.03 & Ic& Liverpool T.\\
2266   & 10 & 16.27$\pm$0.05 & Ic & PROMPT5 & 6898  & 120 & 18.11$\pm$0.03 & Ic& Liverpool T.\\
2355   & 10 & 16.32$\pm$0.05 & Ic & PROMPT5 & 7779  & 180 & 18.14$\pm$0.03 & Ic& Liverpool T.\\
2444   & 10 & 16.22$\pm$0.05 & Ic & PROMPT5 & 8614  & 120 & 18.26$\pm$0.03 & Ic& Liverpool T.\\
2537   & 10 & 16.28$\pm$0.04 & Ic & PROMPT5 & 8942  & 300 & 18.34$\pm$0.02 & Ic& Liverpool T.\\
2626   & 10 & 16.33$\pm$0.05 & Ic & PROMPT5 & 9254  & 300 & 18.41$\pm$0.02 & Ic& Liverpool T.\\
2718   & 10 & 16.46$\pm$0.05 & Ic & PROMPT5 & 38229 & 900 & 20.01$\pm$0.12 & Ic& Liverpool T.\\
2807   & 10 & 16.39$\pm$0.05 & Ic & PROMPT5 & 5451   & 225  & 17.43$\pm$0.04 & Ic & SMARTS\\
2898   & 10 & 16.55$\pm$0.06 & Ic & PROMPT5 & 6030   & 225  & 17.55$\pm$0.04 & Ic & SMARTS\\
2987   & 10 & 16.48$\pm$0.05 & Ic & PROMPT5 & 6881   & 225  & 17.71$\pm$0.04 & Ic & SMARTS\\
3077   & 10 & 16.63$\pm$0.05 & Ic & PROMPT5 & 7452   & 225  & 17.75$\pm$0.04 & Ic & SMARTS\\
3166   & 10 & 16.73$\pm$0.07 & Ic & PROMPT5 & 13631  & 225  & 18.43$\pm$0.04 & Ic & SMARTS\\
3256   & 10 & 16.69$\pm$0.06 & Ic & PROMPT5 & 14203  & 225  & 18.54$\pm$0.04 & Ic & SMARTS\\
3346   & 10 & 16.68$\pm$0.06 & Ic & PROMPT5 & 15054  & 225  & 18.51$\pm$0.04 & Ic & SMARTS\\
3435   & 10 & 16.75$\pm$0.07 & Ic & PROMPT5 & 15624  & 225  & 18.56$\pm$0.04 & Ic & SMARTS\\
3524   & 10 & 16.70$\pm$0.07 & Ic & PROMPT5 & 99214  & 2160 & 20.92$\pm$0.05 & Ic & SMARTS\\
3614   & 10 & 16.80$\pm$0.07 & Ic & PROMPT5 & 106376 & 2160 & 21.07$\pm$0.06 & Ic & SMARTS\\
3704   & 10 & 16.74$\pm$0.06 & Ic & PROMPT5 & 193116 & 2160 & 21.45$\pm$0.07 & Ic & SMARTS\\
3793   & 10 & 16.88$\pm$0.08 & Ic & PROMPT5 & 6480  & 450 & 17.63$\pm$0.04 & Ic & CTIO\\
3898   & 10 & 16.84$\pm$0.06 & Ic & PROMPT5 & 14760 & 450 & 18.47$\pm$0.04 & Ic & CTIO\\
3988   & 10 & 16.90$\pm$0.07 & Ic & PROMPT5 & 60769 & 2640 & 19.84$\pm$0.34 & Ic & MITSuME\\ 
4078   & 10 & 17.01$\pm$0.08 & Ic & PROMPT5 & 30079 & 360 & 19.55$\pm$ 0.01 & i & CQUEAN\\
4167   & 10 & 16.95$\pm$0.07 & Ic & PROMPT5 & 3126  & 177 & 17.07$\pm$0.03 & i & GROND \\
4259   & 10 & 17.11$\pm$0.08 & Ic & PROMPT5 & 3728  & 348 & 17.30$\pm$0.03 & i & GROND \\
5354   & 10 & 17.39$\pm$0.04 & Ic & PROMPT5 & 4508  & 347 & 17.59$\pm$0.03 & i & GROND \\
6093   & 10 & 17.61$\pm$0.05 & Ic & PROMPT5 & 5820  & 871 & 17.92$\pm$0.03 & i & GROND \\
7049   & 10 & 17.67$\pm$0.04 & Ic & PROMPT5 & 7651  & 871 & 18.23$\pm$0.03 & i & GROND \\
8378   & 10 & 17.89$\pm$0.04 & Ic & PROMPT5 & 9479  & 866 & 18.52$\pm$0.03 & i & GROND \\
9698   & 10 & 18.07$\pm$0.04 & Ic & PROMPT5 & 11305 & 873 & 18.67$\pm$0.03 & i & GROND \\
11029  & 10 & 18.17$\pm$0.05 & Ic & PROMPT5 & 13134 & 867 & 18.83$\pm$0.03 & i & GROND \\
12372  & 10 & 18.28$\pm$0.05 & Ic & PROMPT5 & 14959 & 866 & 18.96$\pm$0.03 & i & GROND \\
14394  & 10 & 18.40$\pm$0.04 & Ic & PROMPT5 & 807  &  2 & 15.53$\pm$0.09 & C & Watcher\\
16927  & 10 & 18.58$\pm$0.04 & Ic & PROMPT5 & 810  &  2 & 15.34$\pm$0.09 & C & Watcher\\
19540  & 10 & 18.54$\pm$0.04 & Ic & PROMPT5 & 814  &  2 & 15.48$\pm$0.08 & C & Watcher\\
22342  & 10 & 18.61$\pm$0.04 & Ic & PROMPT5 & 1091 & 30 & 15.93$\pm$0.58 & C & Watcher\\
24369  & 10 & 18.68$\pm$0.05 & Ic & PROMPT5 & 1387 & 30 & 15.94$\pm$0.36 & C & Watcher\\
100848 & 10 & 20.33$\pm$0.12 & Ic & PROMPT5 & 1568 & 30 & 16.45$\pm$0.64 & C & Watcher\\
4425 & 3600 & 17.12$\pm$0.03 & Ic & IAC80 & & & & & \\
4809 & 3600 & 17.21$\pm$0.03 & Ic & IAC80& & & & & \\
5183 & 3600 & 17.30$\pm$0.03 & Ic & IAC80& & & & & \\
\hline \hline
\end{longtable}
}
\end{longtab}

\end{document}